\documentstyle[12pt,amsfonts]{article}

\renewcommand{\L}{{\cal L}}

\newcommand{\be}{\begin{enumerate}}
\newcommand{\ee}{\end{enumerate}}
\newcommand{\ba}{\begin{array}}
\newcommand{\ea}{\end{array}}
\newcommand{\beq}{\begin{equation}}
\newcommand{\eeq}{\end{equation}}
\newcommand{\bqa}{\begin{eqnarray}}
\newcommand{\eqa}{\end{eqnarray}}
\newcommand{\bqas}{\begin{eqnarray*}}
\newcommand{\eqas}{\end{eqnarray*}}

\begin{document}

\newtheorem{defi}{Definition}[section]
\newtheorem{lem}[defi]{Lemma}
\newtheorem{prop}[defi]{Proposition}
\newtheorem{theo}[defi]{Theorem}
\newtheorem{rem}[defi]{Remark}
\newtheorem{cor}[defi]{Corollary}

\newcommand{\qed}{\hfill $\Box$\vspace{.5cm}\medskip}

%\begin{document}

\title{Study of quasi-integrable and non-holonomic deformation of equations in the NLS and DNLS hierarchy}

\author {Kumar Abhinav$^1$\footnote{E-mail: {\tt kumar.abhinav@bilkent.edu.tr}}, Partha Guha$^{2,3}$\footnote{E-mail: {\tt partha@bose.res.in}} and Indranil Mukherjee$^4$\footnote{E-mail: {\tt indranil.m11@gmail.com}}
\and
\\$^1$Department of Physics, Bilkent University\\ 06800 \c{C}ankaya, Ankara, Turkey\\
\and
$^2$S. N. Bose National Centre for Basic Sciences,\\ JD Block, Sector III, Salt Lake, Kolkata - 700106,  India \\ \and
$^3$IHES, Le Bois-Marie 35, Route de Chartres\\ 91440, Bures-sur-Yvette France\\
\and
$^4$ School of Management and Sciences,\\ Maulana Abul Kalam Azad University of Technology,\\ West Bengal, BF 142, Sector I, Salt Lake, Kolkata-700064, India\\
}
\maketitle

%\hspace{1.20 in}

%\vspace{.15 in}

\abstract{The hierarchy of equations belonging to two different but related integrable systems, the Nonlinear Schr\"odinger and its derivative variant, DNLS are subjected to two distinct deformation procedures, viz. quasi-integrable deformation $(QID)$ that generally do not preserve the integrability, only asymptotically integrable, and non-holonomic deformation $(NHD)$ that does. QID is carried out generically for the NLS hierarchy while for the DNLS hierarchy, it is first done on the Kaup-Newell system followed by other members of the family. No QI anomaly is observed at the level of EOMs which suggests that at that level the QID may be identified as some integrable deformation. NHD is applied to the NLS hierarchy generally as well as with the specific focus on the NLS equation itself and the coupled KdV type NLS equation. For the DNLS hierarchy, the Kaup-Newell(KN) and Chen-Lee-Liu (CLL) equations are deformed non-holonomically and subsequently, different aspects of the results are discussed.}

\bigskip
{\bf Mathematics Subject Classifications (2010)}: 35Q55, 37K10, 37K30
\bigskip

{\bf Keywords and Keyphrases}. Nonlinear Schr\"odinger equation, Derivative nonlinear  Schr\"odinger equation,  Quasi-integrable deformation, Non-holonomic deformation, hierarchy.

\bigskip

\section{Introduction}

Completely integrable systems have many important and diverse physical applications such as in water waves, plasma physics, field theory and nonlinear optics \cite{Das}. The standard procedure to study the integrable models is by using the Lax pair, exploiting the zero curvature condition. Systems are considered to be integrable if they contain infinitely many conserved quantities that give rise to the stability of the soliton solutions. These constants of motion delineate the system dynamics, allowing them to be solved by the method of Inverse Scattering Transform (IST) with appropriate variables \cite{AbC,AKNS,Z1}. Another interesting feature of integrable hierarchies is the fact that they possess a local bi-hamiltonian structure \cite{FLT,Magri}. It is well-known that starting from a suitably chosen spectral problem, one can set up a hierarchy of nonlinear evolution equations. One of the important challenges in the study of integrable systems is to determine new such system s which are associated with nonlinear evolution equations of physical significance. \\

The Nonlinear Schr\"odinger (NLS) equation, in one space and one time $(1+1)$ dimensions, is a very well known integrable PDE. It also incorporates semi-classical solitonic solutions, that are physically realizable, and reflect a high degree of symmetry. The latter property corresponds to the infinitely many conserved quantities. There are different variants of the NLS equation such as the coupled KdV type NLS, the generalized NLS, the Kundu-Eckhaus equation, the dimensionless vector NLS etc \cite{BrM,PEZ,Kundu1,CE}. The derivative NLS (DNLS) equation is another celebrated system, its different examples being the Kaup-Newell (KN)\cite{KN}, the Chen-Lee-Liu (CLL) \cite{CLL} and Gerdjikov-Ivanov (GI) equations \cite{GI}. \\

The concept of complete integrability is difficult to establish in case of field-theoretical models because of their infinite number of degrees of freedom. Real physical systems are definitely non-integrable; however, the importance of integrable models in the purview of such systems stems from the fact that the study of continuous physical systems as slightly deformed integrable models is of significant interest. It was recently shown that the sine-Gordon model can be deformed as an approximate system, giving rise to a finite number of conserved quantities \cite{f1,New01}. Some non-integrable models have been shown to possess soliton-like configurations and display properties not significantly different from that of solitons in integrable models, examples being the Ward modified chiral models and the baby Skyrme models with many potentials \cite{O2}. \\

The preceding discussion suggests that we may extend our reasoning beyond integrability and introduce the concept of quasi $(almost)$-integrability. This was precisely the approach of Ferreira {\it et. al.} \cite{O2,1} who considered the modified NLS potential of the form $V ~ (|\psi|^2)^{2+\varepsilon}$, with $\varepsilon$ being a perturbation parameter, and proved that such models possess an infinite number of quasi-conserved charges. Exact dark and bright soliton configurations of QI NLS system \cite{New1,New2} have also been obtained, the latter possessing infinite towers of exactly conserved charges, bringing the system back closer to integrability. QI deformation has also been studied in supersymmetric SG models \cite{Own}. \\

Another situation of current interest is the non-holonomic deformation of integrable systems in which the system is perturbed in such a way that under suitable differential constraints on the perturbing function, the system maintains its integrability \cite{NHD1}. It was shown by Karasu-Kalkani {\it et. al.} \cite{KK} that the integrable sixth order KdV equation represented the non-holonomic deformation of the KdV equation preserving its integrability and generating an integrable hierarchy. The terminology \emph{non-holonomic deformation} (NHD) was used by Kuperschmidt \cite{Kup}. In Ref. \cite{Kundu3} a matrix Lax pair, the N-soliton solution using IST as well as a two-fold integrable hierarchy were obtained by for the non-holonomic deformation of the KdV equation. The work was extended in \cite{KSN} to include the NHD of both KdV and mKdV equations as well as their symmetries, hierarchies and integrability. While studies on the non-holonomic deformation of DNLS and Lenells-Fokas equations were carried out in \cite{Kundu4}, NHD of generalized KdV type equations was discussed in \cite{Gu4}, wherein a geometric angle was provided into the KdV6 equation. In this work, Kirrilov's theory of co-adjoint representation of the Virasoro algebra was used to generate a large class of KdV6 type equations equivalent to the original equation. It was further shown that the Adler-Kostant-Symes approach provided a geometric formalism to obtain non-holonomic deformed integrable systems. NHD for the coupled KdV system was thereby generated. In \cite{Gu5} the author extended Kupershmidt's infinite-dimensional construction  to generate NHD of a wide class of coupled KdV systems, all of which follow from the Euler-Poincar\'e-Suslov flows. \\

\smallskip

The purpose of the present work is to study the behavior of equations in the NLS and DNLS hierarchies when subject to nonholonomic as well as quasi-integrable deformations narrated above. The prior preserves integrability whereas the latter preserves the same in a loose sense; having an infinite number of charges which are asymptotically conserved in
the scattering of soliton-like solutions. These `quasi-integrable charges' are not conserved in time and 
they do vary considerably during the scattering process. However the values coincide with the scattering
with the values they had before. Conservation properties of these QI systems are demonstrated mostly
via numerical methods \cite{f1,New01,O2,1,New1,New2,FerreiraKdV} for the lower order hierarchical equations.
On the other hand nonholonomically deformed systems remain completely integrable \cite{Kup,Kundu3,KSN,Kundu4,Gu4,Gu5}.
The fact that the deformation is applied to the temporal Lax component automatically preserves the
scattering data of the undeformed system \cite{Kundu3,KSN,Kundu4}. The corresponding deformation functions are
exclusively position-dependent, making the final system conservative given the original one being integrable
subjected to higher order constraints. This fact automatically identifies these nonholonomic deformations
as semiholonomic,  which are affine in velocities \cite{NHD1}. The stress in the present work is on the
detailed analysis of these two class of deformations in case of NLS hierarchies. There is no attempt
of comparison between the two given their distinct integrability properties (asymptotic vs higher-order
constraints). However, the study of these two deformations of a particular class of systems in a way extends
their integrability structure, leading to even higher order derivative systems with particular asymptotic
behaviors. We adopt the NLS and DNLS hierarchies for this purpose as their integrability structures are
well-documented. Further, the explicit demonstration of QID and NHD are commonly realized for the lower-order
members of these two hierarchies \cite{f1,1,New1,New2,Kundu4,Own2}. We try to realize the QID and NHD genealogies
of these two hierarchies which, to the best of our knowledge, has not been done before.\\

The paper is organized as follows. Section 2 introduces the NLS hierarchy and points to some specific equations therein. This is followed by a thorough analysis of the quasi-integrable deformation of the equations in the NLS hierarchy. NHD is considered next, first in a generalized format, followed by referencing particular equations of the hierarchy. Section 3 repeats this exercise in respect of the DNLS hierarchy. Section 4 lists some general conclusions and indicates how the work may be extended in future.

\section{The NLS hierarchy}

In this hierarchy, the space (L) and time (M) parts f the Lax pair are respectively given by,

\bqa
&&L=\left( \begin{array}{cc}
\matrix   -i\lambda   &  q \\
      r  &    i\lambda
\end{array} \right)\nonumber\\
{\rm and}\quad&& M = \sum_{m=0}^{n} \lambda^{n-m}\left( \begin{array}{cc}
\matrix   a_m   &  b_m \\
      c_m  &    -a_m
\end{array} \right).\label{E01a}  \eqa
In the above, $a_m$, $b_m$ and $c_m$ are connected through the recurrence relations: \\

\beq \begin{array}{cc}
                                                        a_{mx}=qc_m-rb_m \\
                                                        b_{mx}=-2i b_{m+1}-2qa_m    \\
                                                        c_{mx}=2i c_{m+1}+2ra_m
                                                        \end{array},\label{E02} \eeq
with equations of motion at ${\cal O}\left(\lambda^0\right)$ of spectral space spanned by $\lambda$ as:

\beq
q_t=b_{n,x}+2qa_n(\equiv-2ib_{n+1})\footnote{$m\leq n$.}\quad{\rm and}\quad r_t=c_{n,x}-2ra_n(\equiv2ic_{n+1})
\eeq
Equations $(2)$ are obtained by solving the adjoint representation of the spectral problem or the "stationary" equation $M_x = [L, M]$ while equations $(3)$ derive from the zero curvature condition $L_t - M_x + [L, M] = 0$\\

Some specific values of $a_m$, $b_m$ and $c_m$ are as follows:

\beq \begin{array}{cc}
                                                        a_0=\alpha \hspace{2mm}{\rm a~constant}, \hspace{4mm}
                                                        b_0=0,  \hspace{4mm}
                                                        c_0=0;
                                                        \end{array}\label{E04}  \eeq

\beq \begin{array}{cc}
                                                        b_1=i\alpha q, \hspace{4mm}
                                                        c_1=i\alpha r,    \hspace{4mm}
                                                        a_1=0;
                                                        \end{array}\label{E04}  \eeq

                                                        \beq \begin{array}{cc}
                                                        b_2=-\frac{\alpha}{2} q_x,  \hspace{4mm}
                                                        c_2=\frac{\alpha}{2} r_x,    \hspace{4mm}
                                                        a_2=\frac{\alpha}{2} qr;
                                                        \end{array}\label{E05}  \eeq

                                                         \beq \begin{array}{cc}
                                                        b_3=\frac{i\alpha}{2} (-\frac{1}{2}q_{xx}+q^2 r),  \\
                                                        c_3=\frac{i\alpha}{2} (-\frac{1}{2}r_{xx}+q r^2),  \\
                                                        a_3=\frac{i\alpha}{4} (r q_x-q r_x);
                                                        \end{array}\label{E06}  \eeq

                                                         \beq \begin{array}{cc}
                                                        b_4=\frac{\alpha}{8} (q_{xxx}-6q q_x r),   \\
                                                        c_4=\frac{\alpha}{8} (-r_{xxx}+6q r r_x),   \\
                                                        a_4=\frac{\alpha}{8}(q^2r^2+r_xq_x-qr_{xx}-rq_{xx});\\
                                                        \end{array}\label{E07}  \eeq  \\
                                                        and so on.   \\

The stationary equation $M_x = [L, M]$ can be rewritten as ,
$$(\ M)_{x}-[L,M]=\sum_{m=0}^{n}\lambda^{n-m}\left( \begin{array}{cc}
\matrix{   a_{mx}-qc_m+rb_m   &  b_{mx}+2i\lambda b_m + 2q a_m \\
      c_{mx}-2i\lambda c_m - 2r a_m  &    -(a_{mx}-q c_m+r b_m)}
\end{array} \right),$$
which upon using the recurrence relations and simplifying reduces to,

\beq  \left( \begin{array}{cc}
\matrix   0   &  -2ib_{n+1} \\
      2ic_{n+1}  &    0
\end{array} \right). \eeq

This leads to the NLS hierarchy of equations,

\beq  \begin{array}{lc}
q_t=-2ib_{n+1}  \\
r_t=2ic_{n+1}
\end{array}.\label{N09a}  \eeq
Successive equations of the hierarchy can be generated by putting $n=1,2,3 $ etc. \\
Putting $n=2$, we obtain,
\beq  \begin{array}{lc}
q_t=-2ib_{3} \\
r_t=2ic_{3}
\end{array}.  \eeq

Using the values of $b_3$ and $c_3$  from equation (6) we get,
\beq  \begin{array}{lc}
q_t=\alpha(-\frac{1}{2} q_{xx}+q^2 r) \\
r_t=\alpha(\frac{1}{2} r_{xx}-q r^2),
\end{array}  \eeq
which constitute a system of NLS equations.  \\
Setting $n=3$ leads to,
\beq  \begin{array}{lc}
q_t=i\alpha (-\frac{1}{4} q_{xxx}+\frac{3}{2} q q_x r)  \\
r_t=i\alpha(-\frac{1}{4} r_{xxx}+\frac{3}{2} q r r_x)
\end{array},\label{N012a}  \eeq
which are a pair  of coupled KdV type NLS equations.

\subsection{Quasi-integrable deformation of NLS hierarchy}
The Hamiltonians corresponding to the NLS hierarchy are,

\bqa
&&{\cal H}_1=\int_x\,qr,\nonumber\\
&&{\cal H}_2=\int_x\left(rq_x-qr_x\right),\nonumber\\
&&{\cal H}_3=\frac{1}{2}\int_x\left(q_xr_x+q^2r^2\right),\nonumber\\
&&{\cal H}_4=\int_x\left(qr_{xxx}-3q^2rr_x\right),\nonumber\\
\vdots\label{N01}
\eqa
The corresponding Lax pair, which can be re-expressed as,

\beq
L=-i\lambda\sigma_3+q\sigma_++r\sigma_-,\qquad M=\sum_{m=0}^n\lambda^{n-m}\left(a_m\sigma_3+b_m\sigma_++c_m\sigma_-\right),\label{N02}
\eeq
leads to the zero-curvature condition:

\bqa
&&\left[q_t-\sum_m\lambda^{n-m}\left(b_{m,x}+2a_mq+2i\lambda b_m\right)\right]\sigma_++\left[r_t-\sum_m\lambda^{n-m}\left(c_{m,x}-2ra_m-2i\lambda c_m\right)\right]\sigma_-\nonumber\\
&&=\sum_m\lambda^{n-m}\left(a_{m,x}+rb_m-qc_m\right)\sigma_3,\label{N03}
\eqa
from where the EOM results at ${\cal O}\left(\lambda^0\right)$ and consistency conditions of Eq. \ref{E02} at
${\cal O}\left(\lambda^m\right),~m\leq n$. It is to be noted that the RHS above vanishes by virtue of the first of Eq.s
\ref{E02}, so does the ${\cal O}\left(\lambda^{n\neq 0}\right)$ contributions of the coefficients on the LHS from the
second and third ones of the same set of equations, leaving out the EOMs at ${\cal O}\left(\lambda^0\right)$. On
comparing the expressions for the coefficients in Eq.s \ref{E04}, \ref{E05}, \ref{E06} and \ref{E07} to the expressions
for the Hamiltonians in Eq. \ref{N01}, it is easy to see that,

\beq
b_m=\beta_m\frac{\delta H_m}{\delta r}\quad{\rm and}\quad c_m=\gamma_m \frac{\delta H_m}{\delta q}, \label{N04}
\eeq
where $\beta_m$ and $\gamma_m$ are suitable constants with no sum intended over $m$.

\paragraph*{}Incorporating deformations of the systems in terms of the Hamiltonians, the curvature is expressed as,

\bqa
&&F_{tx}=\left[q_t-\sum_m\lambda^{n-m}\left\{\left(\beta_m\frac{\delta H_m}{\delta r}\right)_x+2a_mq+2i\lambda\beta_m\frac{\delta H_m}{\delta r}\right\}\right]\sigma_+\nonumber\\
&&\qquad+\left[r_t-\sum_m\lambda^{n-m}\left\{\left(\gamma_m\frac{\delta H_m}{\delta q}\right)_x-2ra_m-2i\lambda\gamma_m \frac{\delta H_m}{\delta q}\right\}\right]\sigma_-\nonumber\\
&&\qquad-\sum_m\lambda^{n-m}\left(a_{m,x}+r\beta_m\frac{\delta H_m}{\delta r}-q\gamma_m \frac{\delta H_m}{\delta q}\right)\sigma_3.\label{N05}
\eqa
In general, due to the deformations of Eq.s \ref{N04}, this curvature does not vanish. However, one can always consider
the system, with first two coefficients vanishing, that leads to two deformed EOMs with time-evolution. This is allowed as
one can consider suitable $q$ and $r$ dependence of the Hamiltonians $H_m$s for a given set of $a_m$s. However, once that
is done, it is no longer necessary that the third coefficient in above vanishes, which is aptly identified as the {\it
anomaly}. This particular system is the quasi-deformed NLS hierarchy. Of course a different choice could have been made
with the last expression vanishing, but that would necessarily have meant sacrificing the time-evolution equations for at least one variable between $q$ and $r$. Therefore, from this Quasi-Integrable (QI) mechanism, the set of equations are,

\bqa
&&q_t=\sum_m\lambda^{n-m}\left\{\left(\beta_m\frac{\delta H_m}{\delta r}\right)_x+2a_mq+2i\lambda\beta_m\frac{\delta H_m}{\delta r}\right\},\nonumber\\
&&r_t=\sum_m\lambda^{n-m}\left\{\left(\gamma_m\frac{\delta H_m}{\delta q}\right)_x-2ra_m-2i\lambda\gamma_m \frac{\delta H_m}{\delta q}\right\}\nonumber\\
&&{\rm with~non-vanishing~anomalies:}\nonumber\\
&&{\cal X}_m:=q\gamma_m \frac{\delta H_m}{\delta q}-a_{m,x}-r\beta_m\frac{\delta H_m}{\delta r}.\label{N05a}
\eqa
The ${\cal O}\left(\lambda^0\right)$ anomaly contribution, consistent with time evolution of the system, is,

\beq
{\cal X}_n=q\gamma_n \frac{\delta H_n}{\delta q}-a_{n,x}-r\beta_n\frac{\delta H_n}{\delta r},\label{N06}
\eeq
with higher order contributions accommodated by corresponding deformed versions of Eq.s \ref{E02}.%, which are now {\it non-equalities}.

\paragraph*{}Abelianization can be implemented at this point through suitable $sl(2,c)$ rotation, essentially yielding
the on-shell {\it non}-zero-curvature condition \cite{1},

\beq
\tilde{F}_{tx}\equiv{\cal X}_n\sigma_3.\label{N07}
\eeq
To this end, we choose the gauge operator,

\beq
\tilde{g}=\exp\left(\frac{i}{2}\varphi b^0\right),\label{N07}
\eeq
over the following representation of $sl(2,c)$ algebra:

\bqa
&&b^j=\lambda^j\sigma_3,\qquad F_1^j=\frac{\lambda^j}{2}\left(\kappa\sigma_+-\sigma_-\right), \qquad F_2^j=\frac{\lambda^j}{2}\left(\kappa\sigma_++\sigma_-\right),\quad\kappa\in{\mathbb R};\nonumber\\
&&\left[b^j,b^k\right]=0,\qquad \left[b^j,F_{1,2}^k\right]=F_{2,1}^{j+k},\qquad \left[F_1^j,F_2^k\right]=\frac{\kappa}{2} b^{j+k}.\label{N08}
\eqa
With the identification,

\beq
\exp(2i\varphi)=-\kappa\frac{r}{q},\label{N09}
\eeq
this leads to,

\bqa
&&\tilde{L}=\tilde{g}L\tilde{g}^{-1}+\tilde{g}_x\tilde{g}^{-1}=-ib^1+\frac{i}{2}\varphi_xb^0+2i\sqrt{\frac{qr}{\kappa}}F_1^0,\nonumber\\
&&\tilde{M}=\frac{i}{2}\varphi_tb^0+\sum_{m=0}^n\Big\{a_mb^{n-m}+\left(\frac{1}{\kappa}e^{i\varphi}b_m-e^{-i\varphi}c_m\right)F_1^{n-m}\nonumber\\
&&\qquad\quad+\left(\frac{1}{\kappa}e^{i\varphi}b_m+e^{-i\varphi}c_m\right)F_2^{n-m}\Big\}.\label{N010}
\eqa
As EOMs are utilized, coefficients of $\sigma_\pm$ in Eq. \ref{N05}
can be equated to zero as the deformed EOMs, validating Eq. \ref{N07}, as the $\sigma_3\equiv b^0$ component remains the
same for $\tilde{F}_{tx}\to\tilde{g}F_{tx}\tilde{g}^{-1}$ for $\tilde{g}$ of Eq. \ref{N07}. Therefore:

\beq
\tilde{F}_{tx}\equiv\sum_m\lambda^{n-m}{\cal X}_mb^0.\label{N010a}
\eeq

\paragraph*{}Through {\it another} gauge transformation with respect to,

\beq
\bar{g}=\exp\left(\sum_{j=1}^\infty{\cal F}^{-j}\right),\qquad{\cal F}^{-j}=\xi_1^{-j}F_1^{-j}+\xi_2^{-j}F_2^{-j}\label{N011}
\eeq
an Abelian sub-algebra representation,

\beq
\bar{L}=-ib_1+\sum_{j=0}^\infty {\cal L}_0^{-j}b^{-j},\label{N012}
\eeq
by choosing $\xi_{1,2}^{-j}$s judiciously such that $\bar{L}$ does not depend on $F_{1,2}^{-j}$. The other Lax component
acquires the general expression:

\beq
\bar{M}\equiv\sum_{m=0}^n\lambda^{n-m}a_m\sigma_3+\sum_{j=0}^\infty\left[{\cal M}_0^{-j}b^{-j}+{\cal M}_1^{-j}F_1^{-j}+{\cal M}_2^{-j}F_2^{-j}\right],\label{N013}
\eeq
where ${\cal M}_{1,2}^{-j}$s contain sums containing $b_m$s and $c_m$s. However, their exact forms are not relevant for
calculating the QI anomalies.

\paragraph*{}From Eq. \ref{N07}, the final curvature takes the form,

\beq
\bar{F}_{tx}:=\bar{L}_t-\bar{M}_x+\left[\bar{L},\bar{M}\right]\equiv{\cal X}_n\bar{g}b^0\bar{g}^{-1}:={\cal X}_n\sum_{j=0}^\infty\left[\alpha_0^{-j}b^{-j}+\alpha_1^{-j}F_1^{-j}+\alpha_2^{-j}F_2^{-j}\right],\label{N014}
\eeq
at the lowest spectral order,

\beq
{\cal L}^{-j}_{0,t}-{\cal M}_{0,x}^{-j}={\cal X}_n\alpha_0^{-j},\label{N015}
\eeq
leading to the anomalous charge conservation law,

\beq
\frac{d}{dt}Q^j=\Gamma^j;\quad{\rm where}\quad Q^j=\int_x\,{\cal L}^{-j}\quad{\rm and}\quad \Gamma^j=\int_x\,{\cal X}_n\alpha_0^{-j}.\label{N016}
\eeq
Following Ferreira {\it et. al.}'s treatment \cite{1} of utilizing the two ${\Bbb Z}_2$ transformations available in the
system, namely the $sl(2,c)$ automorphism and space-time parity, the possible reduction of the parent algebra into Image and
Kernel subspaces, allows to show that $\alpha_0^{-j}$s are parity-even. This additionally requires that $q$ and $r$ are
so chosen that $\varphi$ is odd under parity. For those members of NLS hierarchy for which the {\it same} choice of $q$
and $r$ yields parity-odd ${\cal X}_n$ (for example, the standard NLSE), $\Gamma^j$ vanishes asymptotically, ensuring
quasi-integrability.

\paragraph*{}As a short summary of definite parity evaluation of $\alpha_0^{-j}$s, let us introduce the $Sl(2,c)$
automorphism operator ${\frak A}$ and space-time parity operator ${\frak P}$ with actions:

\bqa
&&{\frak A}(b^n)=-b^n,\qquad{\frak A}(F_1^n)=-F_1^n,\qquad{\frak A}(F_2^n)=F_2^n;\nonumber\\
&&{\frak P}:\quad(\tilde{x},\tilde{t})\rightarrow(-\tilde{x},-\tilde{t}),\qquad \tilde{x}=x-x_0,\quad\tilde{t}=t-t_0.\label{N025}
\eqa
Here $(x_0,t_0)$ is any arbitrary origin. $b^n$s define the Kernel subspace, separating it from the Image one:

\beq
{\cal G}={\rm Im}+{\rm Ker};\qquad[b^n,{\rm Ker}]=0,\quad[b^n,{\cal G}]={\rm Im}.\label{N026}
\eeq
Thus $\bar{L}$ in Eq. \ref{N012} lies in the Kernel subspace. As $\bar{L}$ is effected neither by the hierarchy order, nor
by the QID, the constraints satisfied by $\xi_{1,2}^{-j}$s are exactly like those evaluated by
Ferreira {\it et. al.} \cite{1}. On considering different spectral order contributions to $\bar{L}$, we obtain:

\bqa
&&\bar{L}^{(1)}=-ib^1,\nonumber\\
&&\bar{L}^{(0)}=i\left[b_1,{\cal F}^{-1}\right]+\tilde{L}^{(0)},\quad\tilde{L}^{(0)}:=\frac{i}{2}\varphi_xb^0+2i\sqrt{\frac{qr}{\kappa}}F_1^0,\nonumber\\
&&\bar{L}^{(-1)}=i\left[b^1,{\cal F}^{-2}\right]+\left[{\cal F}^{-1},\tilde{L}^{(0)}\right]-\frac{i}{2!}\left[{\cal F}^{-1},\left[{\cal F}^{-1},b^1\right]\right]+{\cal F}^{-1}_x,\nonumber\\
&&\vdots\label{N027}
\eqa
Now considering definite and same parity for $q$ and $r$, as $\varphi$ is parity-odd, it is easy to see from the first
of Eq.s \ref{N010} that,

\beq
\Omega\tilde{L}\equiv-\tilde{L},\qquad\Omega:={\frak A}{\frak P},\label{N028}
\eeq
true for {\it each} part of $\tilde{L}$. Then, from Eq.s \ref{N027},

\bqa
&&(1+\Omega)\bar{L}^{(1)}=0,\nonumber\\
&&(1+\Omega)\bar{L}^{(0)}\equiv i\left[b_1,(1-\Omega){\cal F}^{-1}\right],\nonumber\\
&&\Omega\bar{L}^{(-1)}\equiv-i\left[b_1,\Omega{\cal F}^{-2}\right]-\left[\Omega{\cal F}^{-1},\tilde{L}^{(0)}\right]+\frac{i}{2!}\left[\Omega{\cal F}^{-1},\left[\Omega{\cal F}^{-1},b^1\right]\right]-{\cal F}^{-1}_x,\nonumber\\
\vdots\label{N029}
\eqa
From the second equation above, as all spectral order contribution to $\bar{L}$ are odd under $\Omega$-operation (Eq.
\ref{N012}), the LHS vanishes. So,

\beq
{\rm either}\quad (1-\Omega){\cal F}^{-1}\in{\rm Ker} \quad{\rm or}\quad (1-\Omega){\cal F}^{-1}=0,\label{N030}
\eeq
where the second one conclusion is true by definition. Then, from the third of Eq.s \ref{N029},

\bqa
&&(1+\Omega)\bar{L}^{(-1)}\equiv i\left[b_1,(1-\Omega){\cal F}^{-2}\right],\nonumber\\
or,&&(1-\Omega){\cal F}^{-2}=0.\label{N031}
\eqa
Recursively in this way, one finds $(1-\Omega){\cal F}^{-j}=0$ and therefore,

\beq
(1-\Omega)g=0.\label{N032}
\eeq
Now on considering the {\it Killing form} of the $sl(2)$ loop algebra:

\beq
T_r\left(b^nb^m\right)=\frac{1}{2}\delta_{n,-m},\qquad T_r\left(b^nF_{1,2}^m\right)=0,\qquad T_r\left(\star\right):=-\frac{i}{2\pi}\oint\frac{d\lambda}{\lambda}t_r\left(\star\right),\label{N033}
\eeq
with the second trace over matrices, from Eq. \ref{N014},

\beq
\alpha_0^{-j}=2T_r\left(\bar{g}b^0\bar{g}^{-1}b^n\right)=2T_r\left({\frak A}(\bar{g})b^0{\frak A}(\bar{g}^{-1})b^n\right),\label{N034}
\eeq
as the Killing form is invariant under automorphism and $b^n$s are odd under the same. Therefore,

\beq
{\frak P}(\alpha_0^{-j})\equiv 2T_r\left(\Omega(\bar{g})b^0\Omega(\bar{g}^{-1})b^n\right)=2T_r\left(\bar{g}b^0\bar{g}^{-1}b^n\right)\equiv\alpha_0^{-j}.\label{N035}
\eeq
Thus, following the last of Eq.s \ref{N016}, for the anomaly ${\cal X}_n$ being {\it parity-odd}, we will have
quasi-conservation: $\Gamma^j=0$.

\paragraph*{The particular anomalies:}From the expressions of the Hamiltonian in Eq.s \ref{N01} and the definition of
anomaly in the last of Eq.s \ref{N05a}, we have,

\bqa
&&{\cal X}_1=(\gamma_1-\beta_1)qr,\nonumber\\
&&{\cal X}_2=-\frac{\alpha}{2}(qr)_x-2(\gamma_2qr_x+\beta_2rq_x),\nonumber\\
&&{\cal X}_3=2(\gamma_3+\beta_3)q^2r^2+\beta_3rq_{xx}-\gamma_3qr_{xx}-i\frac{\alpha}{4}(rq_{xx}-qr_{xx}),\nonumber\\
&&{\cal X}_4=q\gamma_4(r_{xxx}-6qrr_x)+r\beta_4(q_{xxx}-6rqq_x)-\frac{\alpha}{8}(q^2r^2+r_xq_x-qr_{xx}-rq_{xx}),\nonumber\\
&&\vdots,\label{N036}
\eqa
where the expressions from Eq.s \ref{E04}-\ref{E07} have been utilized. As $n=2$ gives the usual
NLSE, with $q$ and $r$ being parity-even, for $\gamma_2=\beta_2$, we have ${\cal X}_2\propto(qr)_x$ which is parity-odd,
therefore serving the purpose of quasi-integrability (Ref. \cite{1}). A similar result is obtained for $n=4$.
However, for $n=1,3$, the corresponding anomaly is parity-even. By simple observation of power of the variables in the
NLS hierarchy, it is clear that only {\it even} ordered systems can be quasi-integrable.

\paragraph*{Explicit coefficients:}For the sake of completion, we express the relations satisfied by the $\bar{g}$
coefficients in order to have $\bar{L}$ in the Kernel subspace as follows:

\bqa
&&{\cal O}(0):\quad \xi_1^{-1}=0,\quad\xi_2^{-1}=-2\sqrt{qr/\kappa},\nonumber\\
&&{\cal O}(1):\quad \xi_1^{-2}=-2i\left(\sqrt{qr/\kappa}\right)_x,\quad\xi_2^{-2}=-\varphi_x\sqrt{qr/\kappa},\nonumber\\
&&{\cal O}(2):\quad \xi_1^{-3}=-i\varphi_{xx}\sqrt{qr/\kappa}-2i\varphi_x\left(\sqrt{qr/\kappa}\right)_x,\nonumber\\
&&\qquad\qquad \xi_2^{-3}=2\left(\sqrt{qr/\kappa}\right)_{xx}-\frac{1}{2}(\varphi_x)^2\sqrt{qr/\kappa}-\frac{4}{3}\kappa\left(qr/\kappa\right)^{3/2},\nonumber\\
&&{\cal O}(3):\quad \xi_1^{-4}=2i\left(\sqrt{qr/\kappa}\right)_{xxx}-4i\kappa\sqrt{qr/\kappa}\left(\sqrt{qr/\kappa}\right)_x+\frac{3}{2}\varphi_x\varphi_{xx}\sqrt{qr/\kappa}\nonumber\\
&&\qquad\qquad\qquad+\frac{3}{2}(\varphi_x)^2\left(\sqrt{qr/\kappa}\right)_x,\nonumber\\
&&\qquad\qquad \xi_2^{-4}=\varphi_{xxx}\sqrt{qr/\kappa}+3\varphi_{xx}\left(\sqrt{qr/\kappa}\right)_x+3\varphi_x\left(\sqrt{qr/\kappa}\right)_{xx}-\frac{1}{4}(\varphi_x)^3\sqrt{qr/\kappa}\nonumber\\
&&\qquad\qquad\qquad-\frac{10}{3}\kappa\varphi_x(qr/\kappa)^{3/2},\nonumber\\
&&\vdots.\label{N048}
\eqa
Thus, given the solution is known, the coefficients can be evaluated in principle. In accord with these constraints, the
`anomaly coefficients' $\alpha_0^{n-m,-j}$, for a given set of $(n,m)$ are,

\bqa
&&\alpha_0^{n-m,0}=1,\quad\alpha_0^{n-m,-1}=0,\quad\alpha_0^{n-m,-2}=\frac{\kappa}{4}\left\{\left(\xi_2^{-1}\right)^2-\left(\xi_1^{-1}\right)^2\right\}\equiv qr,\nonumber\\
&&\alpha_0^{n-m,-3}=\frac{\kappa}{2}\left(\xi_2^{-1}\xi_2^{-2}-\xi_1^{-1}\xi_1^{-2}\right)\equiv\varphi_xqr,\nonumber\\
&&\alpha_0^{n-m,-4}=\frac{\kappa}{4}\left\{2\xi_2^{-1}\xi_2^{-3}-2\xi_1^{-1}\xi_1^{-3}+\left(\xi_2^{-2}\right)^2-\left(\xi_1^{-2}\right)^2\right\}+\frac{\kappa^2}{96}\left\{\left(\xi_1^{-1}\right)^2-\left(\xi_2^{-1}\right)^2\right\}^2\nonumber\\
&&\qquad\qquad\equiv\frac{3}{2}(qr)^2+\frac{3}{4}(\varphi_x)^2qr+\left\{\left(\sqrt{qr}\right)_x\right\}^2-2\sqrt{qr}\left(\sqrt{qr}\right)_{xx},\nonumber\\
&&\vdots.\label{N049}
\eqa
From Eq.s \ref{N016}, $\Gamma^1=0$, thereby imposing quasi-integrability for given $(n,m)$. For the anomaly ${\cal X}_n$
being a total derivative, $\Gamma^0=0$ too. The charges are given as space-integral of the coefficients of $\bar{L}$:

\bqa
&&{\cal L}_0^1=-i,\quad{\cal L}_0^0=\frac{i}{2}\varphi_x,\quad{\cal L}_0^{-1}=iqr,\quad{\cal L}_0^{-2}=\frac{i}{2}\varphi_xqr,\nonumber\\
&&{\cal L}_0^{-3}=-i\left\{\left(\sqrt{qr}\right)_x\right\}^2+\frac{i}{4}\left(\varphi_x\right)^2qr+\frac{i}{2}(qr)^2,\cdots\label{N057}
\eqa
Among them, following Eq.s \ref{N049}, the particular ones are conserved, namely, ${\cal L}_0^{-1}$.

\subsection{Non-holonomic deformation of NLS hierarchy}
The non-holonomic deformation of the NLS hierarchy starts with the deformation of the temporal Lax component in Eq. \ref{E01a} by an
amount,

\beq
\delta M=\sum_l\lambda^l{\cal G}_l,\qquad {\cal G}_l={\frak a}_l\sigma_3+{\frak b}_l\sigma_++{\frak c}_l\sigma_-.\label{N037}
\eeq
This effectively amounts to the modifications:

\bqa
&&a_m\to a_m^d=a_m+{\frak a}_{n-m},\quad b_m\to b_m^d=b_m+{\frak b}_{n-m},\quad c_m\to c_m^d=c_m+{\frak c}_{n-m},\nonumber\\
&&{\rm for}\quad l=n-m,\label{N038}
\eqa
with other values of $l$ contributing through the remainder of $\delta M$, leading to the characteristic non-holonomic
constraints. Eq.s \ref{E02}-\ref{E07} are altered accordingly. From Eq. \ref{N03} the dual set of equations appear as,

\bqa
&&q_t=\sum_m\lambda^{n-m}\left(b^d_{m,x}+2a^d_mq+2i\lambda b^d_m\right),\nonumber\\
&&r_t=\sum_m\lambda^{n-m}\left(c^d_{m,x}-2ra^d_m-2i\lambda c^d_m\right),\label{N039}
\eqa
accompanied by the set of non-holonomic constraints:

\bqa
&&{\frak a}_{l,x}=a{\frak c}_l-r{\frak b}_l,\nonumber\\
&&{\frak b}_{l,x}+2i{\frak b}_{l-1}+2q{\frak a}_l=0,\nonumber\\
&&{\frak c}_{l,x}=2i{\frak c}_{l-1}+2r{\frak a}_l,\qquad{\rm where},\quad l\neq n-m.\label{N040}
\eqa
Finally, Eq.s \ref{N09a}-\ref{N012a} are extended by the deformation parameters. \\

We illustrate the non-holonomic deformation by considering the specific example of NLSE given by equation (11) with the choice $\alpha = - i$. The spatial and temporal components of the Lax pair of this equation are given by,
\bqa   &&\begin{array}{lc}
L=-i \lambda \sigma_3 +q \sigma_+ + r \sigma_-,
\end{array}\nonumber\\
&&\begin{array}{lc}
M_{\rm Original}=-i \lambda^2 \sigma_3 + \lambda(q \sigma_+ + r\sigma_-) + {-(\frac{i}{2})qr\sigma_3 + (\frac{i}{2}q_x)\sigma_+ - (\frac{i}{2}r_x)\sigma_-},
\end{array}   \eqa
where the subscript indicates the time component prior to deformation. To obtain the deformation, let us introduce,
\beq   \begin{array}{lc}
M_{\rm Deformed}=\frac{i}{2}\sum_{n=1}\lambda^{-n} G^{(n)},
\end{array}   \eeq
where,
\beq   \begin{array}{lc}
G^{(n)}=a^n\sigma_3+g^n_+\sigma_++g^n_-\sigma_-.
\end{array}   \eeq
The spectral order $n$ can take any integer value which directly or indirectly determines the order of the constraint
equations, and thereby, that of the resultant hierarchy itself, as will be seen. The time part of the Lax pair takes the
form
\beq   \begin{array}{lc}
\tilde{M}=M_{\rm Original} + M_{\rm Deformed}
\end{array}   \eeq   
The zero curvature condition used with $L$ and $\tilde{M}$ shows that while the positive powers of $\lambda$ are trivially satisfied, the zeroth power (or the $\lambda$ free term) leads to the perturbed dynamical systems (equations), while the negative powers of $\lambda$ give rise to the differential constraints. The deformed pair of the NLS equations are given by,
\beq   \begin{array}{lc}
q_t-\frac{i}{2}q_{xx}+iq^2r=-g_1,\quad g_1=g^1_+,
\end{array}   \eeq

\beq   \begin{array}{lc}
r_t+\frac{i}{2}r_{xx}-iqr^2=g_2,\quad g_2=g^1_-.
\end{array}   \eeq
Crucially, only the $\lambda^{-1}$ term from $M_{\rm Deformed}$ contributes in deforming the dynamical equation,
whereas all the other values of $n>1$ contribute only to the constraint conditions, and thereby to the hierarchy itself.
Therefore, there is a clear sectioning in the spectral space.\\
In order to elucidate this, we consider the simplest case of the perturbation $M_{\rm Deformed}\equiv\lambda^{-1}G^{(1)}$.
Then, on equating the $\lambda^{-1}$ order coefficients of the generators $\sigma_3$, $\sigma_+$, $\sigma_-$ from the
zero curvature condition successively, we obtain the following individual constraint conditions on the functions $a$, $g_1$ and $g_2$ as,
\beq   \begin{array}{lc}
a_x = q g_2 - r g_1,
\end{array}   \eeq
\beq   \begin{array}{lc}
g_{1x} + 2aq = 0,
\end{array}   \eeq
\beq   \begin{array}{lc}
g_{2x} - 2ar = 0.
\end{array}   \eeq
The foregoing equations can be shown to give rise to the differential constraint:

\beq   \begin{array}{lc}
\hat{L}(g_1, g_2) = r g_{1xx} + q_x g_{2x} + 2qr (q g_2 - r g_1) = 0.
\end{array}   \eeq
On eliminating the deforming functions $g_1$ and $g_2$, we can derive a new higher order equation as,

\beq   \begin{array}{lc}
-r (q_t - \frac{i}{2} q_{xx} + i q^2 r)_{xx} + q_x (r_t + \frac{i}{2} r_{xx} - i q r^2)_x \\ +  2qr [q (r_t + \frac{i}{2} r_{xx} - i q r^2) + r(q_t - \frac{i}{2} q_{xx} + iq^2 r)] = 0,
\end{array}   \eeq
which is a {\it fourth order} equation.\\

Next we consider the contribution up to the second order ($n=2$) deformation of the NLS equation:

\beq   \begin{array}{lc}
M_{\rm Deformed}(\lambda) = \frac{i}{2} (\lambda^{-1} G^{(1)} + \lambda^{-2} G^{(2)}),
\end{array}   \eeq
where the function $G^{(2)}$ is given by,

\beq   \begin{array}{lc}
G^{(2)} = b\sigma_3 + f_1 \sigma_+ + f_2 \sigma_-,
\end{array}   \eeq
and $G^{(1)}$ is already defined in equation $(54)$. The zero-curvature condition is now applied with $L$ as before but
$M_{\rm Deformed}$ as defined in $(63)$. The following results arise:

(i) No change occurs in the deformed NLS equations, as inferred above. From this, it can immediately be concluded that no
contribution from $M_{\rm Deformed}$ with $n>1$ can effect the deformed NLS equation further, as their corresponding
contribution will not occur at the same spectral order ($\lambda^0$) as that equation.

(ii) Picking up the terms in $\lambda^{-1}$ from the zero curvature equation and equating the coefficients of the
generators $\sigma_3$, $\sigma_+$ and $\sigma_-$ successively, we are led to the following individual constraints:

\beq   \begin{array}{lc}
a_x = q g_2 - r g_1,
\end{array}   \eeq

\beq   \begin{array}{lc}
g_{1x} + 2 i f_1 + 2 a q = 0,
\end{array}   \eeq

\beq   \begin{array}{lc}
g_{2x} - 2 i f_2 - 2 a r = 0.
\end{array}   \eeq
The preceding set of equations finally lead to the following differential constraint:

\beq   \begin{array}{lc}
\hat{L}(g_1,g_2)+2i(r f_{1x}-q_x f_2)=0,
\end{array}   \eeq
with,
\beq   \begin{array}{lc}
\hat{L}(g_1,g_2)=r g_{1xx} + g_{2x} q_x + 2qr(q g_2-r g_1).
\end{array}   \eeq
On comparison with Eq.s $58-61$, however, it is observed that the constraint conditions do change due to the $n=2$ contribution, yielding a more elaborate structure.

(iii) The terms in $\lambda^{-2}$ give rise to a second constraint,

\beq   \begin{array}{lc}
\hat{L} (f_1, f_2) = 0,
\end{array}   \eeq
where the functional form of the above expression is already given by $(69)$
while $f_1$, $f_2$ make up the argument in $(70)$.
However, from Eq.s $66-67$, $f_{1,2}$ are first order in derivatives in $g_{1,2}$. Therefore, the elimination of $f_{1,2}$
from Eq. $70$ in terms of $q$ and $r$ will now lead to a {\it fifth order} differential equation, unlike the case with
only $n=1$ perturbation (Eq. $62$).

Thus, this is an example where the perturbed equations are kept the same, but the order of the differential constraint is
increased recursively, thereby creating a new integrable hierarchy for the NLS equation. We may also consider NHD of systems other than the NLS system, with order of NHD focused on highlighting the complete spectral
sector that deforms the dynamics. Any different order extension to $M_{\rm deformed}$ will only build-up the constraint-induced hierarchy.

For the sake of completeness, we now discuss the coupled KdV type NLSE, for which the space and time components of the Lax pair are given by:
\beq   \begin{array}{lc}
L=-i\lambda \sigma_3 + q \sigma_+ + r \sigma_-   \hspace{4mm}  \mbox{and}    \\
M_{\rm Original}=-i\lambda^3 \sigma_3 + \lambda^2(q \sigma_+ + r \sigma_-) + \lambda[(-\frac{i}{2}q r)\sigma_3 + \frac{i}{2}q_x \sigma_+ -\frac{i}{2} r_x \sigma_-] \\
\qquad\qquad\quad+[\frac{1}{4}(r q_x - q r_x)\sigma_3 + (\frac{1}{2}q^2 r - \frac{1}{4}q_{xx})\sigma_+ + (\frac{1}{2}q r^2 - \frac{1}{4}r_{xx})\sigma_-].
\end{array}   \eeq
$M_{\rm Original}$ now includes a term in $\lambda^3$ as compared to $\lambda^2$ in the previous example of the NLS equation. This would lead to a higher order dispersion term. We take $M_{\rm Deformed}$  = $\frac{i}{2}\lambda^{-1}G^{(1)}$
and therefore, $\tilde{M}$ = $M_{\rm Original} + M_{\rm Deformed}$. Then, on using the zero-curvature condition, we arrive at the following deformed equations:

\beq   \begin{array}{lc}
q_t + \frac{1}{4} q_{xxx} - \frac{3}{2} q q_x r = - g_1,
\end{array}   \eeq
and

\beq   \begin{array}{lc}
r_t + \frac{1}{4} r_{xxx} - \frac{3}{2} r r_x q = g_2,
\end{array}   \eeq
along with the differential constraint:

\beq   \begin{array}{lc}
\hat{L} (g_1, g_2) = 0.
\end{array}   \eeq
In this example, the constraint is held fixed at its lowest level, but the order of the NLS equation is increased (terms enter with higher order dispersion) and thus a new integrable hierarchy is formed.

\section{The DNLS hierarchy}
In this case the Lax operator $L$ is taken as,

\beq
\L= \left( \begin{array}{cc}
\matrix   -i\lambda^2 - is   &  \lambda q \\
      \lambda r  &    i\lambda^2+is
\end{array} \right).\label{N041}
 \eeq
The M operator is slightly complicated given by,
\beq
 M ={\tilde{V}}^{(n)} + \triangle_{n},
\eeq
where,
$${\tilde{V}}^{(n)} = (\lambda^{2n+2} V)_{+}, \qquad
V=
\left( \begin{array}{cc}
\matrix   a   &  b \\
      c  &    -a
\end{array} \right),
$$
and the elements $a$, $b$ and $c$ are expanded in negative powers of $\lambda$. In the above $\triangle_{n}$ = $2 \beta a_{2(n+1)} \sigma_3 (0)$, where $\beta$ is a constant. $\sigma_3 (0)$ is the loop algebra generator corresponding to $\lambda=0$.
The coefficients of the elements $a$, $b$ and $c$, when expanded in negative powers of $\lambda$, are governed by the following recurrence relations:

\beq \begin{array}{cc}
a_{mx}=q c_{m+1}-r b_{m+1} \\
b_{mx}=-2i b_{m+2}-2i s b_m -2q a_{m+1}    \\
c_{mx}=2i c_{m+2} + 2i s c_m + 2ra_{m+1}
\end{array},\label{N053}  \eeq
from which it can be shown that,
\beq
a_{(m+1)x}=(r s b_m - q s c_m) - \frac{i}{2} (q c_{mx} + r b_{mx}).
\eeq

In general it is found that:
\beq
a_{2j+1}=0, b_{2j}=0, c_{2j}=0,\label{N051}
\eeq
for $j=0,1,2,3,....$. The zero-curvature equation leads to the following dynamical systems,

\beq  \begin{array}{l}
q_t= b_{(2n+1) x} + i (1+2\beta) q r b_{2n+1} + 2(1+2\beta) q a_{2(n+1)}   \\
r_t = c_{(2n+1) x} - i (1+2\beta) q r c_{2n+1} - (1+2\beta) r a_{2(n+1)}
\end{array}.\label{N054}  \eeq
These represent a coupled system of hierarchy of equations. Here, we have substituted,

\beq
s=\frac{1}{2}(1+2\beta)qr,
\eeq
following the equations of motion from zero-curvature condition, leading to the expression of $(qr)_t$, and
also that at ${\cal O}\left(\lambda^0\right)$, $(1+2\beta)a_{2n+2,x}\equiv-is_t$.

Putting  $n=1$, we obtain,
\beq  \begin{array}{l}
q_t = b_{3x} + i (1+2\beta) q r b_3 + 2(1+2\beta) q a_4   \\
r_t = c_{3x}  - i(1+2\beta) \beta q r c_3 - 2(1+2\beta) r a_4.\label{31}
\end{array}  \eeq
After using  $2s = (1+2\beta) q r$  , we obtain the following expressions for  $b_3, c_3$  and  $a_4$  viz.
\beq   \begin{array}{l}
b_3= i q_x - 2 \beta  q^2 r   \\
c_3= -i r_x - 2 \beta q r^2   \\
a_4= \frac{1}{2} (r q_x - q r_x) + (2 \beta + \frac{1}{4})i q^2 r^2.
\end{array}  \eeq
Hence  Eq. \ref{31}  yields,
\beq  \begin{array}{l}
q_t= i q_{xx} - (4 \beta +1) q^2 r_x -4 \beta q q_x r +\frac{i}{2}(1+ 2\beta) (4 \beta +1) q^3 r^2   \\
r_t= -i r_{xx} - (4 \beta +1) r^2 q_x -4 \beta r r_x q -\frac{i}{2}(1+ 2\beta) (4 \beta +1) q^2 r^3.\label{33}
\end{array}   \eeq
Eq.s  \ref{33} represent coupled Kundu-type systems. 

Several reductions of Eq. \ref{33} are possible. On putting  $\beta = -\frac{1}{2}$, we get,
\beq   \begin{array}{l}
q_t = i q_{xx} + (q^2 r)_x    \\
r_t = -i r_{xx} + (q r^2)_x,
\end{array}   \eeq
which form a coupled Kaup-Newell (KN) system.  $ \beta = -\frac{1}{4}$ leads to,
\beq   \begin{array}{l}
q_t = i q_{xx} + q q_x r    \\
r_t = -i r_{xx} + r r_x q,
\end{array}   \eeq
which is the coupled Chen-Lee-Liu (CLL) system. Finally,  $ \beta =0$  yields,
\beq   \begin{array}{l}
q_t = i q_{xx} - q^2 r_x + \frac{i}{2} q^3 r^2      \\
r_t = -i r_{xx} - r^2 q_x - \frac{i}{2} q^2 r^3,
\end{array}   \eeq
which is a coupled GI system. Putting  $ r = q^*$ in the above system of equations leads to further reductions, {\it e. g.} setting  $ r = q^*$  in  \ref{33} leads to,
\beq
q_t = i q_{xx} - (4 \beta +1) q^2 q_x^* - 4 \beta |q|^2 q_x + i \beta (4 \beta +1) |q|^4 q,
\eeq
which again represents Kundu type equation.

\subsection{Quasi-integrable deformation of DNLS hierarchy}
We consider the set of KN equations,

\beq
q_t = i q_{xx} + (q^2 r)_x   \quad{\rm and}\quad r_t = -i r_{xx} + (q r^2)_x,\label{N017}
\eeq
obtained for $\beta=-1/2$, as the starting-point for the QI deformation of DNLS system. The specific Lax pair for this
system is,

\bqa
&&L=-i\lambda^2\sigma_3+\lambda q\sigma_++\lambda r\sigma_- \quad{\rm and}\nonumber\\
&&M\equiv\left(\lambda^2a_2-2i\lambda^4\right)\sigma_3+\left(\lambda b_3+\lambda^3b_1\right)\sigma_++\left(\lambda c_3+\lambda^3c_1\right)\sigma_-,\label{N018}
\eqa
with,

\beq
a_2=-iqr,\quad b_1=2q, \quad c_1=2r,\quad b_3=iq_x+q^2r \quad{\rm and}\quad c_3=-ir_x+r^2q.\label{N019}
\eeq
This system originates from a Hamiltonian,

\beq
H=\frac{1}{2}\int_x\left(iq_xr-ir_xq-q^2r^2\right),\label{N020}
\eeq
corresponding to the PB structure,

\beq
\left\{q(x),r(y)\right\}=\frac{1}{2}(\partial_x-\partial_y)\delta(x-y),\label{N021}
\eeq
following the usual definition of time-evolution: $\alpha_t=\{\alpha,H\}$.

\paragraph*{}To attain the QI deformation of the system in Eq. \ref{N017}, the following identifications are made:

\beq
E_q:=q_t - i q_{xx} - \left(\frac{\delta H}{\delta r}+iq_{xx}\right)_x=0   \quad{\rm and}\quad E_r:=r_t +i r_{xx} - \left(\frac{\delta H}{\delta q}+ir_{xx}\right)_x=0,\label{N022}
\eeq
corresponding to the coefficients,

\beq
a_2=-\frac{i}{2}\frac{\delta^2H}{\delta q\delta r},\quad b_3=iq_x+\frac{\delta H}{\delta r}+iq_{xx} \quad{\rm and}\quad c_3=-ir_x+\frac{\delta H}{\delta q}+ir_{xx}.\label{N023}
\eeq
The corresponding curvature tensor takes the form,

\bqa
&&F_{tx}\equiv\left\{\lambda E_q+\lambda^3\left(\frac{\delta H}{\delta q}+ir_{xx}-\frac{1}{2}r\frac{\delta^2H}{\delta q\delta r}\right)\right\}\sigma_++\left\{\lambda E_r+\lambda^3\left(\frac{\delta H}{\delta r}+iq_{xx}-\frac{1}{2}q\frac{\delta^2H}{\delta q\delta r}\right)\right\}\sigma_-\nonumber\\
&&\qquad+{\cal X}\sigma_3;\quad{\rm where},\nonumber\\
&&{\cal X}:=\lambda^2\left[\frac{i}{2}\left(\frac{\delta^2H}{\delta q\delta r}\right)_x+q\left(-ir_x+\frac{\delta H}{\delta q}+ir_{xx}\right)-r\left(iq_x+\frac{\delta H}{\delta r}+iq_{xx}\right)\right].\label{N024}
\eqa
Now, if the Hamiltonian $H$ is deformed, Eq.s \ref{N022} will represent the deformed EOMs, whereas ${\cal X}$ will
represent the QI anomaly at a {\it different} spectral order, vanishing identically for the undeformed system.

\paragraph*{}Before embarking on quasi-deformation, a few points are to be taken note of:

\begin{enumerate}
\item The standard Abelianization will be different, as the anomaly ${\cal X}$ is of a different order than the EOMs;
unlike the case of SG, SSG, NLS or KdV systems.
\item However upon Abelianization, for some $\alpha_0^{-j}$ being constant, only ${\cal X}$ will take part in the
conservation expression:
$$\frac{dQ}{dt}=\int_x{\cal X}.$$
So having it as a total derivative or to be parity odd will do. Looking at its expression in Eq.s \ref{N024}, this will
amount to having the term:
$$q\frac{\delta H}{\delta q}-r\frac{\delta H}{\delta r}$$
as a total derivative. Albeit, this is subjected to finding {\it constant} terms as coefficient of a particular order in
the $sl(2)$ gauge rotation of the $F_{tx}$.
\item Alternatively, to utilize {\it parity} and {\it automorphism} properties of the system to establish quasi-integrability,
It should be possible in the same line as Ferreira {\it et. al.} as the spectral order of $\sigma_3$ term is different
that those of $\sigma_\pm$ in the expression for $L$ in Eq.s \ref{N018}.
\end{enumerate}

\paragraph*{}It is to be noticed that the spatial Lax component of the KN system in Eq. \ref{N018} is just $\lambda$
times that for the NLS systems. Therefore, we perform a gauge-rotation by the {\it same} operator as that in Eq.
\ref{N07}, that leads to $\lambda$ times the expression in the first of Eq.s \ref{N010}, {\it i. e.},

\beq
\tilde{L}\equiv-ib^2+\frac{i}{2}\varphi_xb^1+2i\sqrt{\frac{qr}{\kappa}}F_1^1.\label{N042}
\eeq
Here also, there is an underlying $sl(2,c)$ loop algebra, exactly similar to that explained in Eq.s \ref{N08}. From the
above equation, it is clear that all the generic structures regarding the ${\Bbb Z}_2$ symmetry treatment of the system
will pass through, leading to a parity-even counterpart to $\alpha_0^{-j}$ of Eq.s \ref{N016} and \ref{N035}. Therefore,
the only job left is to determine the parity property of the current anomaly term ${\cal X}$, whose expression has already
been determined. For the sake of completeness, we provide the expression for the temporal Lax component below:

\beq
\tilde{M}\equiv\left(\frac{i}{2}\varphi_t+\lambda^2a_2-2i\lambda^4\right)\sigma_3+e^{i\varphi}\left(\lambda b_3+\lambda^3b_1\right)\sigma_++e^{-i\varphi}\left(\lambda c_3+\lambda^3c_1\right)\sigma_-.\label{N043}
\eeq
As both single and double space derivatives appear explicitly only with $q$ and $r$ in Eq.s \ref{N024}, for
definite parity solutions $q$ and $r$, the anomaly cannot have definite parity and thus the corresponding $\Gamma^j$s
do not vanish in general. However for particular form of $H$, the overall ${\cal X}$ can still be odd. Additionally,
if it is a {\it total derivative}, then for particular $\alpha_0^{-j}$s with $\partial_x\alpha_0^{-j}=0$, the corresponding
charges will be conserved, yielding quasi-integrability.

\paragraph*{}In order to determine the coefficients $\alpha_0^{-j}$s, we consider the rotated Lax component,

\beq
\bar{L}=\bar{g}\tilde{L}\bar{g}^{-1}+\bar{g}_x\bar{g}^{-1},\label{N044}
\eeq
and impose the condition that it is confined to the sub-algebra spanned by $b^n$s. The corresponding constraints are:

\bqa
&&{\cal O}(1):\quad\xi_1^{-1}=0,\quad \xi_2^{-1}=-2\sqrt{qr/\kappa},\nonumber\\
&&{\cal O}(0):\quad\xi_1^{-2}=0,\quad\xi_2^{-2}=-\varphi_x\sqrt{qr/\kappa}\nonumber\\
&&{\cal O}(-1):\quad\xi_1^{-3}=-2i\left(\sqrt{qr/\kappa}\right)_x,\quad\xi_2^{-3}=-\frac{4}{3}\kappa\left(qr/\kappa\right)^{3/2}-\frac{1}{2}\left(\varphi_x\right)^2\sqrt{qr/\kappa},\nonumber\\
&&{\cal O}(-2):\quad\xi_{1}^{-4}=-i\varphi_{xx}\sqrt{qr/\kappa}-2i\varphi_x\left(\sqrt{qr/\kappa}\right)_x,\nonumber\\
&&\qquad\qquad\xi_{2}^{-4}=-\frac{1}{4}\left(\varphi_x\right)^3\sqrt{qr/\kappa}-3\kappa\varphi_x\left(qr/\kappa\right)^{3/2},\nonumber\\
&&\vdots.\label{N045}
\eqa
The above relations are to be implied while evaluating the coefficients in the expression of,

\beq
\bar{F}_{tx}=\bar{g}\tilde{F}_{tx}\bar{g}^{-1},\label{N046}
\eeq
leading to the expressions:

\bqa
&&\alpha_0^0=1,\quad\alpha_0^{-1}=0,\quad\alpha_0^{-2}=\frac{\kappa}{4}\left\{\left(\xi_2^{-1}\right)^2-\left(\xi_1^{-1}\right)^2\right\}\equiv qr,\nonumber\\
&&\alpha_0^{-3}=\frac{\kappa}{2}\left(\xi_2^{-1}\xi_2^{-2}-\xi_1^{-1}\xi_1^{-2}\right)\equiv\varphi_xqr,\nonumber\\
&&\alpha_0^{-4}=\frac{\kappa}{4}\left\{2\xi_2^{-1}\xi_2^{-3}-2\xi_1^{-1}\xi_1^{-3}+\left(\xi_2^{-2}\right)^2-\left(\xi_1^{-2}\right)^2\right\}+\frac{\kappa^2}{96}\left\{\left(\xi_1^{-1}\right)^2-\left(\xi_2^{-1}\right)^2\right\}^2\nonumber\\
&&\qquad\equiv\frac{3}{2}(qr)^2+\frac{1}{4}(\varphi_x)^2qr,\nonumber\\
&&\vdots.\label{N047}
\eqa
Therefore, from observation alone, for order $j=1$, the corresponding charge is conserved and for $j=0$, the same is true
given ${\cal X}$ is a total derivative. These criteria may lead to quasi-integrability. The charges are given as
space-integral of the coefficients of $\bar{L}$:

\beq
{\cal L}_0^2=-i,\quad{\cal L}_0^1=\frac{i}{2}\varphi_x,\quad{\cal L}_0^0=iqr,\quad{\cal L}_0^{-1}=\frac{i}{2}\varphi_xqr,\quad{\cal L}_0^{-2}=\frac{i}{4}(\varphi_x)^2qr+\frac{i}{2}(qr)^2,\cdots\label{N056}
\eeq
Among them, following Eq.s \ref{N047}, the particular ones are conserved, namely, ${\cal L}_0^{-1}$.

To study the quasi-integrable deformation for other members of the DNLS hierarchy, the form of the corresponding Hamiltonian is essential. From Ref. \cite{NN1}, the general form of the Hamiltonian for the DNLS hierarchy is given as,

\beq
{\cal H}_j^{\rm D}=\frac{1}{2j}\left\{4a_{2j+2}-rb_{2j+1}-qc_{2j+1}\right\},\quad j\geq1,\label{N050}
\eeq
wherein Eq.s \ref{N051} has been effectively considered. This enables general definitions of the form:
$\gamma_k=\gamma_k\left(\delta{\cal H}_k^{\rm D}/\delta q,\delta{\cal H}_k^{\rm D}/\delta r\right)$, where $\gamma$
stands for $(a,b,c)$, which explicitly are:

\beq
a_{2j+2}=-i\frac{j}{2}\left({\cal H}_j^{\rm D}-r\frac{\delta{\cal H}_j^{\rm D}}{\delta r}-q\frac{\delta{\cal H}_j^{\rm D}}{\delta q}\right),\quad b_{2j+1}=-2j\frac{\delta{\cal H}_j^{\rm D}}{\delta r},\quad c_{2j+1}=-2j\frac{\delta{\cal H}_j^{\rm D}}{\delta q}.\label{N052}
\eeq
This enables us to incorporate QIDs directly at the level of Lax pair coefficients. Then from Eq.s \ref{N054}, the most
general curvature expression for the deformed system can be expressed as,

\bqa
&&F^{\rm D}_{tx}\equiv\Big[\lambda q_t+2\lambda^{2n+2}\sum_j\lambda^{-(2j+1)}j\left(\frac{\delta{\cal H}_j^{\rm D}}{\delta r}\right)_x+4i\lambda^{2n+2}\left\{\lambda^2+\left(\frac{1}{2}+\beta\right)qr\right\}\nonumber\\\nonumber\\
&&\qquad\quad \times\sum_j\lambda^{-(2j+1)}j\frac{\delta{\cal H}_j^{\rm D}}{\delta r}+2i\lambda q\Big\{\lambda^{2n+2}\sum_j\lambda^{-(2j+2)}\frac{j-1}{2}\left({\cal H}_{j-1}^{\rm D}-r\frac{\delta{\cal H}_{j-1}^{\rm D}}{\delta r}-q\frac{\delta{\cal H}_{j-1}^{\rm D}}{\delta q}\right)\nonumber\\
&&\qquad\quad+\beta n\left({\cal H}_{n}^{\rm D}-r\frac{\delta{\cal H}_{n}^{\rm D}}{\delta r}-q\frac{\delta{\cal H}_{n}^{\rm D}}{\delta q}\right)\Big\}\Big]\sigma_+\nonumber\\
&&\qquad+\Big[\lambda r_t+2\lambda^{2n+2}\sum_j\lambda^{-(2j+1)}j\left(\frac{\delta{\cal H}_j^{\rm D}}{\delta q}\right)_x-4i\lambda^{2n+2}\left\{\lambda^2+\left(\frac{1}{2}+\beta\right)qr\right\}\nonumber\\\nonumber\\
&&\qquad\quad \times\sum_j\lambda^{-(2j+1)}j\frac{\delta{\cal H}_j^{\rm D}}{\delta q}-2i\lambda r\Big\{\lambda^{2n+2}\sum_j\lambda^{-(2j+2)}\frac{j-1}{2}\left({\cal H}_{j-1}^{\rm D}-r\frac{\delta{\cal H}_{j-1}^{\rm D}}{\delta r}-q\frac{\delta{\cal H}_{j-1}^{\rm D}}{\delta q}\right)\nonumber\\
&&\qquad\quad+\beta n\left({\cal H}_{n}^{\rm D}-r\frac{\delta{\cal H}_{n}^{\rm D}}{\delta r}-q\frac{\delta{\cal H}_{n}^{\rm D}}{\delta q}\right)\Big\}\Big]\sigma_-\nonumber\\
&&\qquad+\Big[-i\left(\frac{1}{2}+\beta\right)(qr)_t-\lambda^{2n+2}\sum_j\lambda^{-(2j+2)}\frac{j-1}{2}\left({\cal H}_{j-1}^{\rm D}-r\frac{\delta{\cal H}_{j-1}^{\rm D}}{\delta r}-q\frac{\delta{\cal H}_{j-1}^{\rm D}}{\delta q}\right)_x\nonumber\\
&&\qquad\quad+in\beta\left({\cal H}_{n}^{\rm D}-r\frac{\delta{\cal H}_{n}^{\rm D}}{\delta r}-q\frac{\delta{\cal H}_{n}^{\rm D}}{\delta q}\right)_x\nonumber\\
&&\qquad\quad-2\lambda^{2n+3}\left\{q\sum_j\lambda^{-(2j+1)}j\left(\frac{\delta{\cal H}_j^{\rm D}}{\delta q}\right)-r\sum_j\lambda^{-(2j+1)}j\left(\frac{\delta{\cal H}_j^{\rm D}}{\delta r}\right)\right\} \Big]\sigma_3.\label{N055}
\eqa
In the above, at ${\cal O}(\lambda^1)$, the coefficients of $\sigma_\pm$ represents deformed coupled equations of the
hierarchy, while the coefficient of $\sigma_3$ is the anomaly term. It is to be noted that the constraints of Eq.s
\ref{N053} that lead to Eq.s \ref{N051} are {\it still} valid as only the forms of he coefficients $(a,b,c)$ have been
generalized. This essentially ensures that there will be {\it no} anomaly contribution at ${\cal O}(\lambda^1)$ or at
any odd order in general. This is a crucial fact as there will be no QI anomaly at the level of EOMs, and thus
the QID should be identified as some {\it integrable} deformation, and thus can possibly be identified
with some NHD. At other spectral orders, however, there
will be anomaly contributions, which can directly be determined from Eq. \ref{N055}. Then, quasi-conserved charges can be
determined in accord with Eq.s \ref{N047}.

\subsection{Non-holonomic deformation of DNLS hierarchy}
To discuss the non-holonomic deformation of the equations in the DNLS hierarchy, attention is first focused on the Kaup-Newell system with the Lax pair given in Eq.s \ref{N018}. The modified temporal component of the Lax pair is given as
$\tilde{M}$ = $M_{original} + M_{deformed}$, with $M_{original}$ given by Eq.s \ref{N018} and

\beq   \begin{array}{lc}
M_{deformed} = i (G^{(0)} + \lambda^{-1} G^{(1)} + \lambda^{-2} G^{(2)}),
\end{array}   \eeq
where,

\beq   \begin{array}{lc}
G^{(0)} = w \sigma_3 + m_1 \sigma_+ + m_2 \sigma_-,
\end{array}   \eeq
\beq   \begin{array}{lc}
G^{(1)} = a \sigma_3 + g_1 \sigma_+ + g_2 \sigma_-,
\end{array}   \eeq
\beq   \begin{array}{lc}
G^{(2)} = b \sigma_3 + f_1 \sigma_+ + f_2 \sigma_-.
\end{array}   \eeq
Using the zero-curvature relation, we obtain the following deformed equations:

\beq   \begin{array}{lc}
q_t - \frac{i}{2} q_{xx} - \frac{1}{2} (q^2 r)_x + 2 g_1 - 2 i q w = 0,
\end{array}   \eeq
\beq   \begin{array}{lc}
r_t + \frac{i}{2} r_{xx} - \frac{1}{2}(q r^2)_x - 2 g_2 + 2 i r w = 0.
\end{array}   \eeq
Further, we obtain the following conditions on the different components of the deforming functions $G^{(i)}$:
$m_1 = 0$, $m_2 = 0$, $a = 0$, $f_1 = 0$, $f_2 = 0$ and $b_x = 0$ which implies that $b = b(t)$ only.

We are, therefore, left with the following deforming functions:
\beq   \begin{array}{lc}
G^{(0)} = w (x, t) \sigma_3,\\
G^{(1)} = g_1 (x, t) \sigma_+ + g_2 (x, t) \sigma_-,\\
G^{(2)} = b (t) \sigma_3.
\end{array}   \eeq
Moreover, the following constraints are obtained:

\beq   \begin{array}{lc}
g_{1x} = - 2 q (x, t) b (t),\\
g_{2x} =  2 r (x, t) b (t),\\
w_x = q g_2 - r g_1.
\end{array}   \eeq
It is possible to obtain new nonlinear integrable equations by resolving the constraint relations and expressing all the perturbing functions through the basic field variables. To this end, we put, 

\beq   \begin{array}{lc}
q = u_x, \qquad r = v_x,
\end{array}   \eeq
where $u = u (x, t)$ and $v = v (x, t)$.
Equation $(115)$ used in equation $(114)$ allows us to express $g_1$, $g_2$ and w in terms of $b (t)$, u and v only as follows :

\beq   \begin{array}{lc}
g_1 = -2 b (t) u,\\
g_2 = 2 b(t) v,\\
w = 2 b(t) u v + K (t),
\end{array}   \eeq
where K is again a function of t only. On eiminating $g_1$, $g_2$ and w from  equations $(111)$ and $(112)$, we can rewrite the coupled perturbed $(deformed)$ DNLS equations in the following form:

\beq   \begin{array}{lc}
u_{xt} - \frac{i}{2} u_{xxx} - \frac{1}{2} (u_x^2 v_x)_x - 4 u b (t) - 2 i u_x (2 b (t) u v + K (t)) = 0,
\end{array}   \eeq
\beq   \begin{array}{lc}
v_{xt} + \frac{i}{2} v_{xxx} - \frac{1}{2} (u_x v_x^2)_x - 4 v b (t) + 2 i v_x (2 b (t) u v + K (t)) = 0.
\end{array}   \eeq 
These are coupled evolution equations which are non-autonomous with arbitrary time-dependent coefficients $b (t)$ and $K ( t)$. Clearly, no more constraints are left at this stage. Equations $(117)$ and $(118)$ generalize the coupled system of Lenells-Fokas equations by including a nonlinear derivative term as well as a higher order dispersion term.
Also, it is to be noted that in this system both $\lambda^0$ and $\lambda^{-1}$ ({\it i. e.} $n=0,1$) effects the dynamics, whereas $n>2$ contributions only build the hierarchy up.

We now consider the Chen-Lee-Liu (CLL) system for which the Lax pair is given by,

\beq   \begin{array}{lc}
L = \lambda^2 \left(\matrix{    -i & 0 \\ 0 & i  }    \right) + \lambda \left( \matrix { 0 & q \\ r & 0} \right) + \left (\matrix { 0 & 0 \\ 0 & \frac{i}{2} q r} \right),
\end{array}   \eeq
and,
\beq   \begin{array}{lc}
  M = 2 \lambda^4 \left(\matrix {   -i  &  0  \\   0  &  i}  \right)  +  2 \lambda^3 \left(\matrix {   0  &  q  \\   r  &  0}  \right) +  \lambda^2 q r \left(\matrix {   -i  &  0  \\   0  &  i}  \right) +  \lambda \left(\matrix {   0  &  i q_x + \frac{1}{2} q^2 r  \\   - i r_x + \frac{1}{2} q r^2  &  0}  \right)\\
\qquad\quad+ \left(\matrix {   0  &  0  \\   0  &  - \frac{1}{2} (r q_x - r_x q) + \frac{i}{4} r^2 q^2}  \right).
\end{array}   \eeq
We consider,
\beq   \begin{array}{lc}
M_{deformed} = i (G^{(0)} + \lambda^{-1} G^{(1)} + \lambda^{-2} G^{(2)}),
\end{array}   \eeq
with $G^{(0)} = w \sigma_3$, $G^{(1)} = g_1 \sigma_+ + g_2 \sigma_-$, $G^{(2)} = b \sigma_3$ where we have taken the cue from the preceding discussion in choosing the form of the matrices $G^{(0)}$, $G^{(1)}$ and $G^{(2)}$. On taking  $\tilde{M}$ = $M + M_{deformed}$, and imposing the zero-curvature condition, we are led to the following deformed CLL equations:

\beq   \begin{array}{lc}
q_t = i q_{xx} + q q_x r - 2 g_1 + 2 i q w,
\end{array}   \eeq
and,
\beq   \begin{array}{lc}
r_t = - i-r_{xx} + r r_x q + 2 g_2 - 2 i r w.
\end{array}   \eeq
The following differential constraints are also obtained:

\beq   \begin{array}{lc}
i w_x = q g_2 - r g_1,
\end{array}   \eeq
\beq   \begin{array}{lc}
g_{1x} + 2 q b + \frac{i}{2} q r g_1 = 0,
\end{array}   \eeq
and,
\beq   \begin{array}{lc}
g_{2x} - 2 r b - \frac{i}{2} q r g_2 = 0.
\end{array}   \eeq
We further get $b_x = 0$ which implies that b is a function of t only. However, it may be noted that it is not possible in this case to resolve the constraints and express the perturbing functions through the basic field variables by re-defining these variables. This is due to the presence of a nonlinear term in $(125)$ and $(126)$.

\section{Discussion and conclusion}
Two different deformation techniques are considered in this paper, both being applied to equations belonging to the NLS and DNLS families. For the NLS hierarchy, the QI deformation is done in detail yielding the particular anomalies, the explicit coefficients and the like. In case of DNLS, QID is first applied to the Kaup-Newell system and then for other members of the hierarchy leading to the significant observation that there cannot be any QI anomaly at the level of EOMs which means that in this case, the QID may be identified as some integrable deformation. 

NHD is first generically applied to the NLS hierarchy followed by specific cases; first in case of the NLS equation itself and then for the coupled KdV type NLSE. In case of the DNLS hierarchy, NHD is carried out on two different systems, viz. the Kaup-Newell and Chen-Lee-Liu equations. The two different deformation procedures, one exactly preserving integrability (by construction) and the other only asymptotically, applied to two separate hierarchies, demonstrate an extended class of dynamical systems. These deformed systems, both (non-holonomically) integrable or/and quasi-integrable, adds to the known hierarchies as possible dynamical systems which could be of physical interest with possibly new aspects. Simultaneous application of these techniques to other families of integrable systems and/or their supersymmetric generalizations will be the topic for our future investigation.

\paragraph*{{\it Acknowledgement}:}The research of KA is supported by the T\"UBITAK 2216 grant number 21514107-115.02-124285 of the Turkish government. PG would like to express his gratitude to Professor Maxim Kontsevich and other members of the IHES for their warm hospitality.

\end{document}